\documentclass[fleqn,usenatbib]{mnras}

\usepackage{newtxtext,newtxmath}

\usepackage[T1]{fontenc}

\DeclareRobustCommand{\VAN}[3]{#2}
\let\VANthebibliography\thebibliography
\def\thebibliography{\DeclareRobustCommand{\VAN}[3]{##3}\VANthebibliography}

\usepackage{graphicx}	
\usepackage{amsmath}	

\title[Flipped orbit of KELT-19Ab from gravity darkening]{The flipped orbit of KELT-19Ab inferred from the symmetric TESS transit light curves}

\author[Y. Kawai]{
Yugo Kawai,$^{1}$\thanks{yugo6581@g.ecc.u-tokyo.ac.jp}
Norio Narita,$^{2,3,4}$
Akihiko Fukui,$^{2,4}$
Noriharu Watanabe,$^{1}$
Satoshi Inaba$^{5}$
\\
$^{1}$Department of Multi-Disciplinary Sciences, Graduate School of Arts and Sciences, The University of Tokyo, Tokyo 153-8902, Japan\\
$^{2}$Komaba Institute for Science, The University of Tokyo, Tokyo 153-8902, Japan\\
$^{3}$Astrobiology Center, Tokyo 181-8588, Japan\\
$^{4}$Instituto de Astrofisica de Canarias (IAC), 38205 La Laguna, Tenerife, Spain\\
$^{5}$School of International Liberal Studies, Waseda University, Tokyo 169-8050, Japan
}

\date{Accepted XXX. Received YYY; in original form ZZZ}

\pubyear{2015}

\begin{document}
\label{firstpage}
\pagerange{\pageref{firstpage}--\pageref{lastpage}}
\maketitle

\begin{abstract}
Dozens of planets are now discovered with large orbital obliquity, and have become the proof for the dynamical evolution of planetary orbits. In the current samples, there is an apparent clustering of planets around $90^\circ$, and also an absence of planets around $180^\circ$ although the latter is expected by some theories. Statistical extrapolation using Hierarchical Bayesian Analysis have recently refuted the significant clustering around $90^\circ$ and suggested that the distribution may actually be broader. In this work, the symmetric TESS transit light curve of KELT-19Ab is analyzed using gravity darkening to measure its true obliquity. Its large sky projected obliquity $\lambda = -179.7^{\circ+3.7^\circ}_{\,\,-3.8^\circ}$ makes KELT-19Ab the only currently known planet with obliquity potentially close to $180^\circ$. We apply spectroscopic constraints on $v\mathrm{sin}i$ and $\lambda$ as well as theoretical constraints on the limb-darkening coefficients to find that the KELT-19Ab's obliquity is $\psi = 155^{\circ+17^\circ}_{\,\,-21^\circ}$, in favor of a flipped orbit. The result is consistent with the statistically inferred uniformity of obliquity distribution, and also highlights the applicability of the gravity darkening technique to symmetric light curves.
\end{abstract}




\section{Introduction} \label{sec:intro}
In the past decade of exoplanetary research, orbits of many exoplanets have been discovered misaligned with respect to the rotation axes of their host stars. The degree of misalignment between the orbital axis of exoplanets and the spin axis of their host stars $\psi$ (spin-orbit angle or obliquity) has then become an important probe to investigate the dynamical history of these planets. 

In the current samples of obliquity $\psi$, there is an apparent clustering of planets at around $\psi = 90^\circ$. The trend was first mentioned by \cite{Albrecht_2021}, and also presented in \cite{Attia_2023}. However, multiple studies have more recently refuted such a preponderance of polar planets via Hierarchical Bayesian Analysis \citep{Siegel_2023,Dong_2023}, arguing that the observed peak is statistically insignificant. Such a context puts the absence of planets with obliquity close to $180^\circ$ into question. 

\begin{table}
\caption{$\lambda$ and $\psi$ of retrograde planets}
\centering
\begin{tabular}{lll}
   \hline \hline 
   Planet (Reference) & $\lambda\quad(^\circ)$ & $\psi\quad(^\circ)$\\
   \hline 
   HAT-P-7b(1,2,3,4,5,6,7)&$142.0^{\mathrm{+12.0}}_\mathrm{-16.0}$&$97.0^{\mathrm{+8.0}}_\mathrm{-4.0}$\\
   K2-290b,c(8)& $143.0^\pm\mathrm{8.0}$&$124.0\pm\mathrm{6.0}$\\
   WASP-8b(9,10)& $143.0^{\mathrm{+1.5}}_\mathrm{-1.6}$&$118.2^{\mathrm{+3.2}}_\mathrm{-3.0}$\\
   WASP-94Ab(11,12)&$151.0^{\mathrm{+16.0}}_\mathrm{-23.0}$&$116.6^{\mathrm{+9.9}}_\mathrm{-9.1}$\\
   WASP-167b(13,10)&$165.0\pm\mathrm{5.0}$&$123.8^{\mathrm{+11.6}}_\mathrm{-10.6}$\\
   TOI-640b(14)&$184\pm3$&$104\pm2$\\ 
   \hline
  \multicolumn{3}{l}{\footnotesize1 \cite{Narita_2009}; 2 \cite{Winn_2009};}\\
  \multicolumn{3}{l}{\footnotesize3 \cite{Lund_2014}; 4 \cite{Campante_2016};}\\
  \multicolumn{3}{l}{\footnotesize5 \cite{Masuda_2015}; 6 \cite{Albrecht_2012};}\\
  \multicolumn{3}{l}{\footnotesize7 \cite{Benomar_2014}; 8 \cite{Hjorth_2021};}\\  
  \multicolumn{3}{l}{\footnotesize9 \cite{Queloz_2010}; 10 \cite{Bourrier_2017};}\\
  \multicolumn{3}{l}{\footnotesize11 \cite{Neveu-VanMalle_2014};}\\
  \multicolumn{3}{l}{\footnotesize12 \cite{Albrecht_2021}; 13 \cite{Temple_2017}}\\
  \multicolumn{3}{l}{\footnotesize14 \cite{Knudstrup_2023}}\\
  \end{tabular}
\label{lambda_psi}
\end{table}

Previous measurements of $\psi$ for planets with $\lambda$ (the sky-projection of $\psi$) greatly exceeding $\,90^\circ$ show that their orbits are actually closer to polar than flipped\footnote{Throughout the paper, $60^\circ< \psi < 120^\circ$ is referred to as polar, and $\psi > 120^\circ$ as flipped. Retrograde orbit ($\lambda > 90^\circ$) or even large sky-projected obliquity ($\lambda \approx 180^\circ$) does not ensure a flipped orbit, due to the uncertainty in the direction of the stellar rotational axis \citep{Fabrycky_Winn_2009,Xue_2016}.} (or at least on the verge of the two classes), as listed in Table \ref{lambda_psi}, which vaguely hints to the rarity of these "flipped orbits". Theoretically, however, excitation of obliquity to such an extreme have been predicted with variations of Kozai-Lidov mechanism including the concurrent evolution of stellar spin \citep{Storch_Anderson_Lai_2014,Stroch_Lai_2015,Anderson_2016,Storch_2017}. 

\cite{Storch_Anderson_Lai_2014} found that chaotic evolution of stellar spin due to spin-orbit coupling between the precession frequencies of the oblate star and the Kozai frequencies result in extreme obliquities, some of them close to $180^\circ$. The result owes to the fact that obliquity in this context is the actual angle between the stellar spin axis and the planetary orbital axis, instead the conventional angle between the initial and the evolved planetary orbital axis, which is practically limited by the Kozai angle $= 140.8^\circ$ and assumes primordial star-planet alignment. 

Since the direction of stellar spin significantly alters the final obliquity after Kozai-Lidov mechanism, primordial misalignment between the stellar spin axis and the planetary orbital axis can also produce $\psi \approx 180^\circ$ \citep{Anderson_2016, Vick_2022}. Observational evidence for primordial misalignments exist in forms of and inner disks misaligned with respect to the outer disks \citep{Marino_2015,Ansdell_2020}. Misaligned coplanar multi-planet systems \citep{Huber_2013,Hjorth_2021} are also potentially the legacy of primordial misalignment.

Hot Jupiter KELT-19Ab, explored in this paper, has the largest sky projected obliquity $\lambda = -179.7^{\circ+3.7^\circ}_{\,\,-3.8^\circ}$ of currently known exoplanets, and is confirmed to have a retrograde orbit and potentially a flipped one, depending on the stellar inclination. It also belongs to the class of hot Jupiters around host stars above the Kraft break ($T_\mathrm{eff} = 6250$ K). Beyond this point, convective layers in the host become absent and planets are unlikely to be tidally realigned \citep{Winn_2010}. The outcome of any violent orbital evolution in KELT-19Ab is therfore likely retained.

The host is an Am star with $T_\mathrm{eff} = 7500\pm110$ K and $v\mathrm{sin}i = 84.2 \pm 2.0$ km/s \citep{Siverd_2018}, making it an excellent target for analysis with gravity darkening, whose magnitude is more pronounced in hotter and faster rotating stars. The distant companion KELT-19B is a late-G9V/early-K1V star at a projected separation of 160 au \citep{Siverd_2018}, providing the necessary grounds for Kozai-Lidov mechanism to take place.

In this paper, we use gravity darkening to estimate the obliquity of hot Jupiter KELT-19Ab, from its symmetric TESS \citep{Ricker_2015} light curve. We apply spectroscopic constraints derived by \cite{Siverd_2018} to gauge the extent of gravity darkening apriori and also theoretical constraints on limb-darkening coefficients to test the applicability of the gravity darkening method to symmetric transits. Gravity darkening for KELT-19Ab have previously been explored by \cite{Garai_2022} using CHEOPS data, who concluded a non-detection of gravity darkening. They speculated that this was because the degree of anomaly expected for KELT-19Ab's light curve with gravity darkening is about 10 ppm. We argue that the degree of anomaly is 600 ppm instead, and that their non-detection was due to an apparent underestimation of the host's rotational velocity.

The paper is organized as follows. In Section \ref{sec:methods}, we describe the methods used in this work, including data reduction, gravity darkening, and MCMC including the choice of priors. Section \ref{sec:results} and \ref{sec:discussion} provides the results and the discussion respectively.

\section{Methods} \label{sec:methods}
\subsection{TESS photometry}
We extract the light curve for KELT-19Ab from the TESS full-frame images created at 2-minute cadence in Sector 7, which returned four full transits, using the python package $\mathrm{\begin{tt}eleanor\end{tt}}$ \citep{eleanor}. From the options provided with the package, we select PCAFLUX as the flux values, which has the systematics removed via the subtraction of cotrending basis vectors (CBVs) provided by the Science Payload Operations Center (SPOC) pipeline \citep{Jenkins_2016}. CBVs are obtained through Principal Component Analysis and represent systematic trends unique to each operational sector. 

To correct for flux contamination in the light curve, we use information from the TESS target pixel file to remove excess flux from sources other than the target. Specifically, we use the CROWDSAP and FLFRCSAP attributes, which represent the fraction of flux within the aperture that comes from the target and the fraction of target flux captured within the aperture, respectively. We first subtract the excess flux, calculated as $(1 - \mathrm{CROWDSAP})*$ median flux, from the raw flux values. Then, we divide the corrected flux values by FLRCSAP to obtain a final flux that accounts for the effects of crowding.

\subsection{Geometry}

\begin{figure}
    \centering
    \includegraphics[width=\columnwidth]{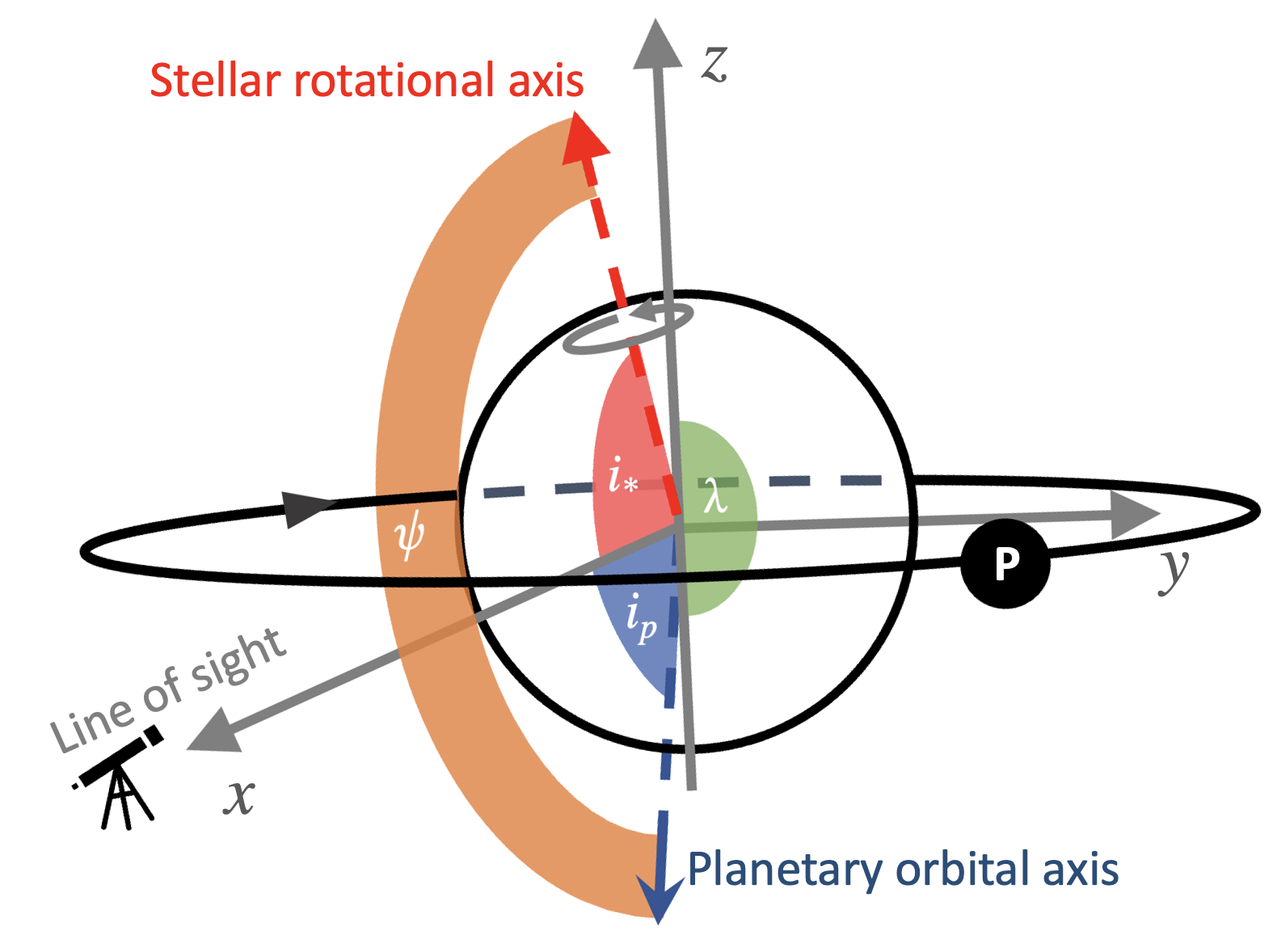}
    \caption{Definition of the relevant angles, true obliquity $\psi$, orbital inclination $i_p$, stellar inclination $i_*$, and sky-projected obliquity $\lambda$.
    }
    \label{fig:geometry}
\end{figure}

The definition for true obliquity $\psi$ used in this paper is given using orbital inclination $i_p$, stellar inclination $i_*$, and sky-projected obliquity $\lambda$ in Figure \ref{fig:geometry}, as well as the equation
\begin{equation}
    \mathrm{cos}\psi=\mathrm{cos}i_\mathrm{p}\mathrm{cos}i_* + \mathrm{sin}i_\mathrm{p}\mathrm{sin}i_*\mathrm{cos}\lambda.
\end{equation}

Here, orbital inclination and stellar inclination are defined as the angle between the planetary orbital axis and stellar rotational axis with the line of sight respectively. 

\begin{figure*}
    \centering
    \includegraphics[width=\textwidth]{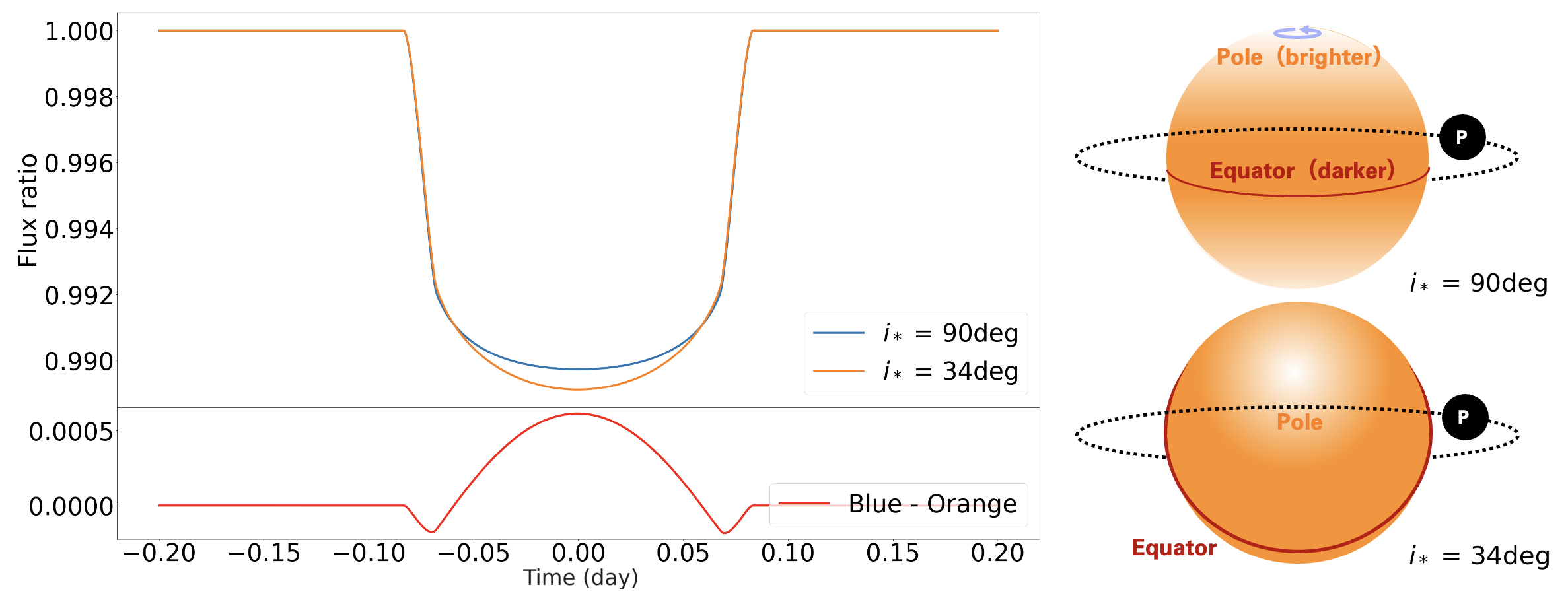}
    \caption{The gravity darkened light curve simulated for KELT-19Ab. The light curve retains its symmetry regardless of $i_*$, owing to sky-projected obliquity $\lambda = -179.7^{\circ+3.7^\circ}_{\,\,-3.8^\circ}$. The blue and orange lines represent the light curve for when $i_* = 90^\circ$ and $i_* = 34^\circ$ respectively. The transit is shallower in the former case, where the planet transits through the darker equator of the gravity darkened host star (Top right). The transit gets progressively deeper towards mid-transit for the latter case, as the planet transits through the brighter region near the pole (Bottom right). The difference in transit depth mid-transit is around 600 ppm.}
    \label{fig:sim_lc}
\end{figure*}

\subsection{Gravity darkening}

Gravity darkening is the latitude dependent flux variation of a star, due to its rapid rotation. This effect causes a transiting planet to occult areas of differing surface brightness at different points in its transit, allowing the obliquity to be constrained by fitting the anomaly in the light curve. The technique has been used to constrain obliquity for planets including KELT-9b \citep{Ahlers_2020}, KOI-89.01 and KOI-89.02 \citep{Ahlers_2015}, KOI-368.01\citep{Zhou_2013}, Kepler-13Ab \citep{Barnes_2011,Masuda_2015}, HAT-P-7b \citep{Masuda_2015}, HAT-P-70b \citep{Zhou_2019}, PTFO~8-8695 \citep{Barnes_2013}, MASCARA-1b \citep{Hooton_2022}, MASCARA-4b \citep{Ahlers_2020b}, WASP-33b \citep{Dholakia_2022}, and WASP-189b \citep{Lendl_2022,Deline_2022}.

Gravity darkening is usually modelled by the von Zeipel theorem \citep{von_Zeipel_1924} as

\begin{equation}
T(\theta) = T_\mathrm{pole}\left(\frac{g(\theta)}{g_\mathrm{pole}}\right) ^\beta,
\label{VonZeipel}
\end{equation}

\noindent where $T(\theta)$ and $g(\theta)$ are the temperature and surface gravity at a given latitude, $T_\mathrm{pole}$ and $g_\mathrm{pole}$ are the temperature and surface gravity at the poles respectively. $\beta$ is the gravity darkening constant, which is equal to 0.25 for a barotropic star in a strict radiative equilibrium. Surface gravity $g(\theta)$ at at a given latitude can be evaluated with the following equation

\begin{equation}
\vec{g}(\theta) = -\frac{GM_*}{r(\theta)^2}\hat{r}+\Omega^2_*\:r(\theta)\mathrm{sin}\theta\:\hat{r}_\bot,
\label{Gravity}
\end{equation}

\noindent where $G$ is the universal gravitational constant, $M_*$ is the stellar mass, $\Omega_*$ is the stellar rotation rate, $r(\theta)$ is the distance from the stellar center to the point of interest, and $r(\theta)\mathrm{sin}\theta$ hence denotes the distance from the stellar rotational axis to the same point. $\hat{r}$ and $\hat{r}_\bot$ are both unit vectors pointing to the point of interest, one from the star's center and the other from the rotational axis. These equations relate the magnitude of centrifugal force at a given latitude to temperature, which will be the lowest at the equator, where the star also becomes the darkest. 

The degree of anomaly is predominantly governed by the stellar effective temperature and rotational velocity from Equation \ref{VonZeipel}, and the general shape of the gravity darkened light curve is governed mainly by the sky-projected obliquity and the stellar inclination. For KELT-19Ab, the light curve retains its symmetry regardless of $i_*$, owing to $\lambda = -179.7^{\circ+3.7^\circ}_{\,\,-3.8^\circ}$. As $i_*$ decreases (approaching pole on), the light curve gets progressively deeper towards mid-transit, as the planet transits through the brighter region near the pole as illustrated in Figure \ref{fig:sim_lc}. The lower limit of $i_* = 34^\circ$ is given by requiring that KELT-19A rotates slower than the empirical limit $v \approx 150
$ km/s for an Am star \citep{Siverd_2018}. The difference in transit depth mid-transit between $i_* = 90^\circ$ and $i_* = 34^\circ$ is around 600 ppm.

On top of the ways in which $\lambda$ and $i_*$ affect the shape of the gravity darkened light curve, it is also worth noting that gravity darkening is usually used in tandem with other methods of constraining $\lambda$, in order to disentangle the degeneracy between the prograde and retrograde solutions. In the hypothetical case that $i_* = 90^\circ$ and the light curve is symmetric, one cannot tell $\psi \approx 0^\circ$ from $\psi \approx 180^\circ$ by gravity darkening alone without prior knowledge of $\lambda$. This is because gravity darkening is a latitude dependent effect which can determine the stellar inclination but not the direction of the host star's spin with respect to the planetary orbit. To get the sense of stellar spin, observations of longitude dependent effects such as the Rossiter-Mclaughlin effect are required.

We use the gravity darkening model implemented in the python package $\mathrm{\begin{tt}PyTransit\end{tt}}$ \citep{pytransit} to estimate the model parameters. For each discretized point on the stellar surface, the model evaluates the temperature based on Eq. \ref{VonZeipel} and derives the flux given an emission spectrum and a response function of the instrument. We use the synthetic spectra from the PHOENIX library \citep{Husser_2013} and the TESS instrument response function provided in \cite{Ricker_2015}. 

\subsection{MCMC parameters and priors}

\begin{table*}
    \caption{List of parameters and their chosen priors}
\centering
\begin{tabular}{llll}
   \hline \hline 
   Symbol & Parameter & GD7400 & GD7600\\
   \hline 
   $\lambda(^\circ)$ & Sky-projected obliquity&\multicolumn{2}{c}{$\mathcal{N}(-179.7,3.8)$}\\
   $v\mathrm{sin}i$(km/s) & Stellar rotational velocity&\multicolumn{2}{c}{$\mathcal{N}(84.8,2)$}\\
   $R_*$($R_\odot$) & Stellar radius&\multicolumn{2}{c}{$\mathcal{N}(1.83,0.099)$}\\
   $R_p/R_*$& Radius ratio& \multicolumn{2}{c}{$\mathcal{U}(0.05,0.25)$}\\
   $b$ & Impact parameter&\multicolumn{2}{c}{$\mathcal{U}(0,1)$}\\
   $\rho_*$ (g/$\textrm{cm}^3$)& Stellar density& \multicolumn{2}{c}{$\mathcal{U}(0,1)$}\\
   $T_\mathrm{eff, pole}$(K)& Stellar effective temperature&7400 (fixed)&7600 (fixed)\\
   $P_\mathrm{orb}$(d)& Orbital period& \multicolumn{2}{c}{4.6117093 (fixed)}\\
   $\beta$& Gravity darkening constant&\multicolumn{2}{c}{0.25 (fixed)}\\
   $u_1$& Quadratic limb-darkening coefficient&$\mathcal{N}(0.2702,0.03)$&$\mathcal{N}(0.2372,0.03)$\\
   $u_2$& Quadratic limb-darkening coefficient&$\mathcal{N}(0.2320,0.03)$&$\mathcal{N}(0.2516,0.03)$\\
   $i_*(^\circ)$&Stellar inclination &\multicolumn{2}{c}{$\mathcal{U}(0,180)$}\\
   \hline\\
  \end{tabular}
\label{Priors}
\end{table*}

The parameters for the gravity darkening model are listed in Table \ref{Priors}, including the sky-projected obliquity $\lambda$, stellar rotational velocity $v\sin i$, stellar radius $R_*$, stellar effective temperature $T_\mathrm{eff}$, gravity darkening exponent $\beta$, stellar inclination $i_*$, as well as regular transit parameters which are radius ratio $R_p/R_*$, impact parameter $b$, stellar density $\rho$, and the two limb-darkening coefficients $u_1,u_2$. Some crucial parameters are also derived from these parameters, which include the stellar rotational period, $P_\mathrm{rot} = 2 \pi R_* \sin i/ v\sin i$, and the stellar oblateness $f$, approximated in the model to be $f = 3\Omega_*^2/8\pi G\rho$. To sample from the posterior distribution, we conduct a Markov chain Monte Carlo (MCMC) run using Python code $\mathrm{\begin{tt}emcee\end{tt}}$, which implements an affine-invariant ensemble sampler \citep{Foreman-Mackey_2013}. We chose 50 walkers, thinning of 10 steps for the entirety of 500,000 steps, and the first 10,000 steps are discarded as burn-in. 

\subsubsection{Specstroscopic constraints}
We apply spectroscopic constraints on the stellar radius, rotational velocity and sky-projected obliquity with normal priors, taking the mean values and uncertainties from \cite{Siverd_2018} as listed in Table \ref{Priors}.

A normal prior on rotational velocity constrains the degree of gravity darkening and hence the degree of anomaly. As discussed by \cite{Barnes_2009}, it then helps distinguish a gravity darkened and symmetric light curve like the one of KELT-19Ab from a regular non gravity darkened light curve. In the exploration of gravity darkening for KELT-19Ab by \cite{Garai_2022}, they have used $\Omega/\Omega_\mathrm{crit} = 0.23$, the ratio of rotational velocity to the break-up velocity, as a constraint instead of $v\mathrm{sin}i$. This seems to be an underestimation, given that ratio of $v\mathrm{sin}i \approx 85$ km/s to the break-up velocity for A-type stars $\approx 250 $ km/s already gives 0.34, which might have resulted in their non-detection of gravity darkening.

A normal prior of $\lambda = -179.7\pm3.8^\circ$ on the sky-projected obliquity constrains the expected shape of the gravity darkened light curve for the reasons discussed in the previous section. It also serves to avoid any nonphysical or discrepant values with the results from Doppler tomography as mentioned in \cite{Masuda_2015}.

\subsubsection{Limb darkening and gravity darkening coefficients}
As obliquity is estimated from subtle anomalies in transit shape, the result of gravity darkening analysis is also sensitive to the choice of limb darkening coefficients \citep{Masuda_2015, Ahlers_2015}. This is especially true for planets with $|\lambda| \approx 0^\circ$ or $180^\circ$, because the transit will always be symmetric and gravity darkening will only be imprinted as the difference in transit depth mid-transit, which is exactly the case for KELT-19Ab. When a gravity darkened light curve retains such U-shaped symmetry, limb-darkening coefficients, if let free, can alter the shape of ingress and egress, causing the problem to be fully degenerate. 

\cite{Siegel_2023} ran a Mote Carlo simulation to show that the aforementioned degeneracy is the cause of an observational bias in gravity darkening. This was done by showing that a regular transit model with free limb-darkening coefficients can provide as good of a fit to any gravity darkened light curves that are U-shaped and symmetric, without the need for a gravity darkening model. Their conclusion was that the method is heavily sensitive to polar or nearly polar planets, which are the limited configurations that potentially produce asymmetric light curves depending on $\lambda$ (See \cite{Barnes_2009} for possible types of asymmetry with gravity darkening).

However, if gravity darkening is expected to occur for a certain system, choice of the gravity darkening model over a regular transit model is not out of the quality of fit, but rather out of the necessity to account for an existing physical phenomenon. In a case a regular transit model is used for gravity darkened light curves that are U-shaped and symmetric, parameters such as radius ratio can be overestimated at the expense of setting limb-darkening coefficients free, because transit depth changes with stellar inclination as seen in Figure \ref{fig:sim_lc}. Therefore, prior constraints on limb-darkening coefficients are necessary.

A common choice is to rely on theoretical values, be it fixing to catalog values as has been done in \cite{Ahlers_2015}, or calculating theoretical priors based on the theoretical stellar intensity profile as in \cite{Deline_2022}\footnote{An alternative approach explored in \cite{Masuda_2015} is to deliberately search for limb-darkening coefficients that returns $\lambda$ in agreement with literature values, without setting prior constraints on $\lambda$ . The discrepancy between those values and theoretical ones was left an open question.}. In this work, we decided to apply normal priors with standard deviation of 0.03 on the limb-darkening coefficients with theoretical values given by \cite{Claret_2017} for TESS bandpass as the mean, given the stellar effective temperature of 7400 K and 7600 K, as well as surface gravity of 4.0 and metallicity of 0. This is based on $T_\mathrm{eff} = 7500\pm110$ K, $\mathrm{log }g_* = 4.127\pm0.029 \,\mathrm{g/cm}^3$ and [Fe/H] $= -0.12\pm0.51$ of KELT-19A from \cite{Siverd_2018}, and on the intent to test the robustness of our results with different choice of limb-darkening coefficients. We hence run two sets of analysis on the light curve, first fixing the stellar effective temperature to 7400 K and then 7600 K. 

The gravity darkening constant is fixed to 0.25. Previous studies have shown that the choice of this constant can lead to conflicting results \cite[e.g.][]{Ahlers_2020b}. To address this, we examined the impact of the gravity darkening exponent on the obliquity of KELT-19Ab by also performing a fit using the theoretical value of around 0.2 proposed by \cite{Claret_2016} for stars with effective temperature, metallicity and surface gravity similar to that of KELT-19A. We find that varying this parameter does not affect the conclusion that KELT-19Ab's orbit could be flipped, and hence leave it to 0.25.

\subsubsection{Remaining parameters}
We let the stellar inclination to vary freely between $0^\circ$ and $180^\circ$, where $0^\circ$ denotes a pole-on configuration. We also apply uniform priors to non-gravity darkening parameters $R_p/R_*$, $b$ and $\rho_*$ in the range specified in Table \ref{Priors}. 

\begin{table*}
    \centering
    \begin{tabular}{l|ccccc}
    $\mathrm{Instrument}$ & \multicolumn{3}{c}{TESS} & KELT& CHEOPS\\\hline
    \textbf{Parameter} & GD7400 & GD7600 &  \cite{Yang_2022}&\cite{Siverd_2018}&\cite{Garai_2022}\\
    \hline
    $(\mathrm{Fitted})$& &\\
    $R_p/R_*$ & $0.09578^{+0.00084}_{-0.00079}$ & $0.09586^{+0.00084}_{-0.00082}$& $0.09576\pm0.00082$ & $0.10713\pm0.00092$ &$0.0985\pm0.0010$\\
    ${\rho_*\mathrm{(g/cm}^3)}$ & $0.615^{+0.057}_{-0.055}$ & $0.610^{+0.058}_{-0.054}$ & - & $0.376^{+0.031}_{-0.027}$&-\\
    $b$ & $0.323^{+0.075}_{-0.099}$ & $0.331^{+0.072}_{-0.096}$& - & $0.601^{+0.026}_{-0.030}$&$0.499\pm0.018$\\
    $i_*(^\circ)$ & $97^{+29}_{-34}$ & $96^{+33}_{-37}$ & - & [33.5,146.5] $1\sigma$ &-\\
    $T_\mathrm{eff,pole}$ & $7400$ (fixed)& $7600$ (fixed)& - & - &-\\
    $P(\mathrm{d})$ & \multicolumn{2}{c}{4.6117093 (fixed)}& \multicolumn{2}{c}{$4.6117093\pm0.0000088$}& $4.6117105\pm0.0000077$\\
    $\beta$ & \multicolumn{2}{c}{0.25 (fixed)} & -&-&-\\
    $\lambda(^\circ)$ & $-179.2\pm3.8$ & $-179.2\pm3.8$& - &$-179.7^{+3.7}_{-3.8}$&-\\
    $v\mathrm{sin}i(\mathrm{km/s})$ & $84.8\pm2.0$& $84.8\pm2.0$& - &$84.8\pm2$ & $86.36\pm0.21$\\
    $R_*(R_\odot)$ & $1.830\pm0.099$& $1.830\pm0.099$&-& $1.830\pm0.099$ &-\\
    $u1$ & $0.256^{+0.023}_{-0.024}$& $0.231^{+0.024}_{-0.023}$ & - & - &-\\
    $u2$ & $0.226^{+0.028}_{-0.028}$& $0.250^{+0.028}_{-0.029}$& - & - &-\\
    $(\mathrm{Derived})$ & &\\
    $a/R_*$ & $8.84^{+0.26}_{-0.27} $& $8.82\pm 0.27$ & $9.04\pm0.19$ & $7.50^{+0.20}_{-0.18}$ & $8.214\pm0.088$\\
    $i_p(^\circ)$ & $87.91^{+0.69}_{-0.57}$& $87.84^{+0.67}_{-0.55}$ &$88.7\pm0.7$& $85.41^{+0.34}_{-0.31}$ &$88.66\pm0.33$\\
    $P_\mathrm{rot}(\mathrm{d})$ & $0.99^{+0.10}_{-0.20}$ &$0.97^{+0.12}_{-0.23}$ &-& -&-\\
    $\psi(^\circ)$ & $157^{+15}_{-20}$ &$155^{+17}_{-21}$ & & [119,180] $1\sigma$ &-\\
    $f$ & $0.016^{+0.009}_{-0.003}$ & $0.016^{+0.012}_{-0.004}$&-&-&-\\
    \hline
    red. $\chi^2$ &1.020 & 1.021 & 1.001 & - & -\\
    \end{tabular}
    \caption{Planetary and stellar parameters obtained from gravity darkening analysis of TESS data assuming $T_\mathrm{eff,pole} = 7400$ K and $T_\mathrm{eff,pole} = 7600$ K as well as from existing literature.}
    \label{Estimated_parameters}
\end{table*}

\section{Results}  \label{sec:results}
\subsection{Obliquity}
\begin{figure*}
    \centering
    \includegraphics[width=1.5\columnwidth]{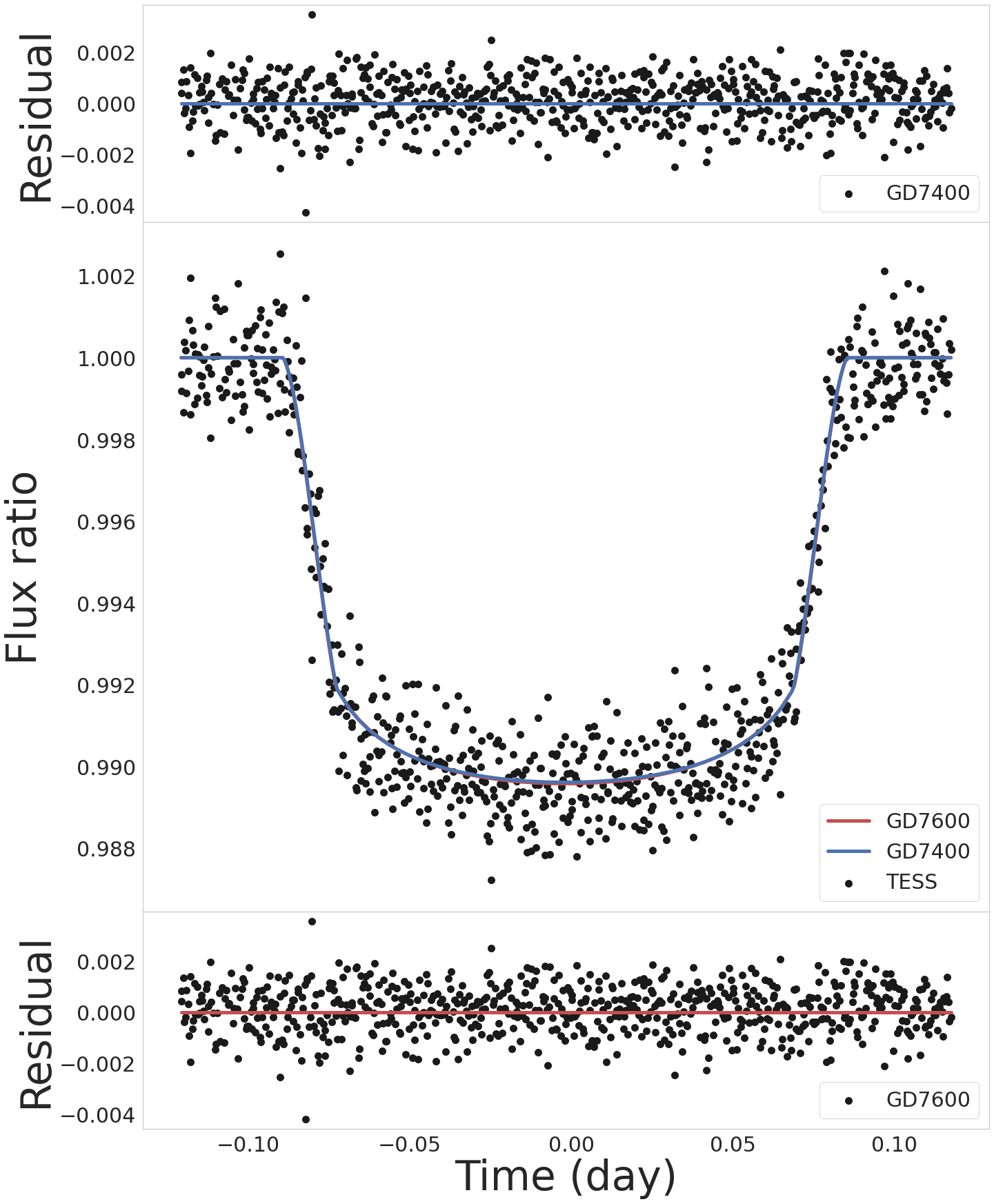}
    \caption{Fit of the two gravity darkening models (GD7400,GD7600) to the phase folded transit light curve from TESS. Both models return very similar fits, resulting in the overlapping lines in the middle plot.}
    \label{fig:fit_result}
\end{figure*}

From the analysis, we obtain that the stellar inclination $i_* = 97^{\circ+29^\circ}_{\,\,-34^\circ}$ and $i_* = 96^{\circ+33^\circ}_{\,\,-37^\circ}$ for the two gravity darkening models (GD7400 and GD7600) respectively. They are fully consistent with $33.5^\circ<i_*<146.5^\circ$ suggested by \cite{Siverd_2018}, assuming that KELT-19A rotates slower than the empirical limit, $v_\mathrm{eq} < 150\,\mathrm{km/s}$. Fit and derived parameters are listed in Table \ref{Estimated_parameters} with the corner plot in Appendix \ref{corner_plots}, and the best-fit models to the data is presented in Figure \ref{fig:fit_result}.

\begin{figure*}
    \centering
    \includegraphics[width=\textwidth]{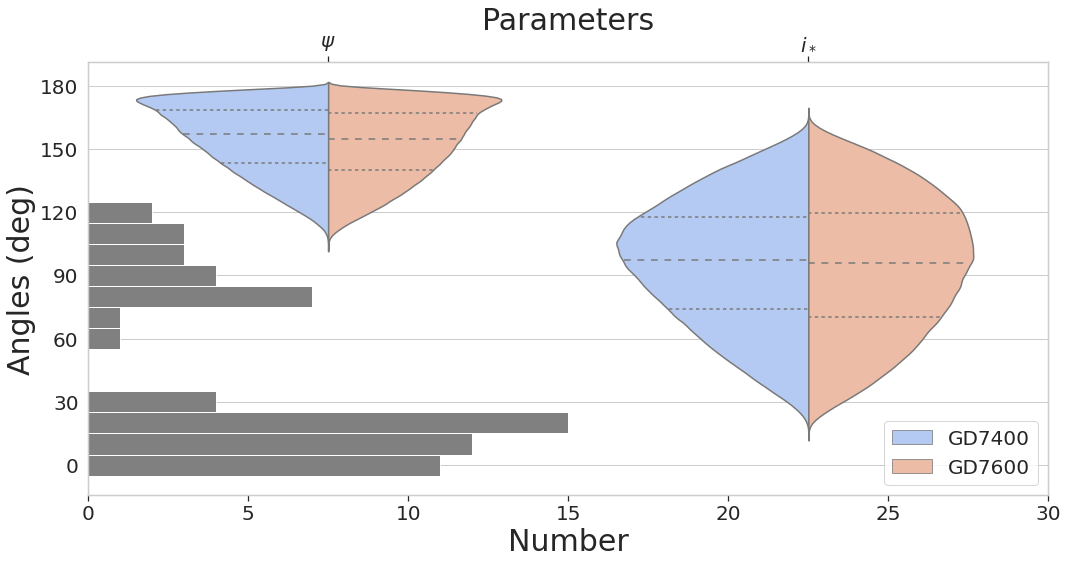}
    \caption{Violin plot showing the posterior distribution for obliquity and stellar inclination derived from the two gravity darkening models. The analysis favors a flipped orbit, and KELT-19Ab likely has the largest obliquity among currently known exoplanets. The over-plotted histogram shows the obliquity of currently known exoplanets taken from \citet{Albrecht_2022} with the count on the bottom x-axis.}
    \label{fig:posterior_plot}
\end{figure*}

The obtained stellar inclination results in $\psi = 157^{\circ+15^\circ}_{\,\,-20^\circ}$ with GD7400 and  $\psi = 155^{\circ+17^\circ}_{\,\,-21^\circ}$, with GD7600, both of which place stronger constraints than the $1\sigma$ constraint of $119^\circ<\psi<180^\circ$ in \cite{Siverd_2018}. At 2 $\sigma$, the constraint with the two models becomes $121^\circ<\psi<177^\circ$ and $119^\circ<\psi<177^\circ$ respectively. The posteriors for both the obliquity and stellar inclination are shown in Figure \ref{fig:posterior_plot}. The result is in favor of a flipped orbit, and KELT-19Ab likely has the largest obliquity among currently known exoplanets, when compared to $124^\circ \pm 6.0^\circ$ of K-290b and c \citep{Hjorth_2021}. 

\subsection{Radius ratio, impact parameter and semi-major axis}

For KELT-19Ab, there are discrepancies in the reported values of the radius ratio, impact parameter and semi-major axis between observations by KELT \citep{Siverd_2018}, CHEOPS \citep{Garai_2022} and TESS \citep{Yang_2022} as listed in Table \ref{Estimated_parameters}. The values obtained in this work for these parameters is consistent with the analysis of \cite{Yang_2022} within $1 \sigma$. \cite{Yang_2022}, who explored the role of atmosphere in this discrepancy but reached an inconclusive result, mentioned that a transit fit with inclination and semi-major axis fixed to KELT values provides a fit with slightly larger reduced $\chi^2$ but still consistent with TESS data, although not the preferred solution. It is therefore likely that the discrepancy is due to some degeneracy.

\cite{Barnes_2009} showed that when $\psi$ and $|\lambda| = 0^\circ$ or $180^\circ$ (i.e. the light curve is symmetric and U-shaped), the radius ratio of a Jupiter sized planet around an gravity darkened Altair (A7V star) with $b = 0.9$ can be overestimated by almost $30\%$ with a regular transit model. This is because in such a gravity darkened system, larger impact parameter causes the planet to transit nearer to the brighter pole, resulting in larger transit depth. Overestimation is less pronounced at lower impact parameters, where increased projected area around the oblate equator makes up for the area being darker. Such an overestimation might be able to explain the trend of larger impact parameter coupled with larger radius ratio seen in Table \ref{Estimated_parameters}. If we then assumed larger estimates of impact parameter in previous literature as the ground truth and perform a fit with gravity darkening, radius ratio should become smaller than in literature.

When impact parameter is fixed to value in \cite{Siverd_2018}, we obtain $0.09870^{+0.00073}_{-0.00084}$ with GD7400 and $0.09863^{+0.00075}_{-0.00087}$ with GD7600, which are both smaller than the values in \cite{Siverd_2018} by 6.7 $\sigma$. The values, however, is marginally smaller but consistent with the $0.09940\pm0.00048$ obtained by \cite{Yang_2022} who did the same analysis using TESS light curve and a quadratic model. When fixed to values from \cite{Garai_2022}, we obtain $0.09743^{+0.00067}_{-0.00076}$ and $0.09741^{+0.00070}_{-0.00080}$ respectively, which are slightly smaller than \cite{Garai_2022}, but still consistent within 1$\sigma$. It is therefore difficult to conclude that gravity darkening is the lone culprit of the reported discrepancies, although it must be playing a role in making the problem degenerate.

It should also be noted, however, that the said overestimation of radius ratio happens only if the larger impact parameters results in occultation of the stellar surface at higher latitude. If the orbit was completely polar ($\psi = 90^\circ$), larger impact parameter will cause the occultation at lower latitude, which must instead lead to an underestimation of radius ratio \citep{Barnes_2009}. This means that the discrepancies can in principle be used to bolster the conclusion about the orbit of planets with symmetric gravity darkened light curves.

\section{Discussion} \label{sec:discussion}
\subsection{Possibility for stronger constraints}
As evident from the posterior distribution, the obtained precision in $\psi$ is limited by the precision in $i_*$, which is one of the primary parameters fitted with the gravity darkening model. Improved observational precision will therefore result in stronger constraints on $i_*$ and thus $\psi$. With double the number of data points, we obtain the photometric precision per two-minute-binned data of $\sim~600$ ppm, which is tantamount to the expected difference in transit depth for the polar and flipped orbit scenario. KELT-19Ab hence remains as an important follow up target for TESS.

Comparing the transit depth over different wavelengths can in principle constrain obliquity, although the required precision is much higher. The degree of gravity darkening varies over different wavelengths, and stronger gravity darkening is expected at shorter wavelengths, resulting in more pronounced pole to equator contrast \citep{Barnes_2009}. When comparing the TESS and CHEOPS passband for example, KELT-19Ab's transit is deeper on the order of $10^1$ ppm with TESS than with CHEOPS, if the planet transits through the darker equator. The transit will instead be deeper on the order of $10^2$ ppm if the planet transits through the pole. Observations at longer wavelengths where limb-darkening is reduced might also be helpful, but this comes at the expense of reduced gravity darkening altogether.

We also consider the chance that changes in impact parameter $b$ or sky-projected obliquity $\lambda$ over time can tell a polar orbit from a flipped one, as the rate of nodal precession is faster for planets on polar orbits than a flipped one. \citep[e.g.][]{Watanabe_2022}. The two parameters are expressed as equations of time as follows.

\begin{equation}
    b(t) = \frac{a}{R_*}(\mathrm{cos}\psi\mathrm{cos}i_*+\mathrm{sin}\psi\mathrm{sin}i_*\mathrm{cos}\theta(t))
\end{equation}
\begin{equation}
    \mathrm{tan}\lambda(t) = \frac{\mathrm{sin}\psi\mathrm{sin}\theta(t)}{\mathrm{sin}\psi\mathrm{cos}i_*\mathrm{cos}\theta(t)-\mathrm{cos}\psi\mathrm{sin}i_*}.
\end{equation}

Here, $\theta(t)$ is the nodal angle, which expresses how much the orbital axis has precessed around the stellar spin axis, and is expressed,

\begin{equation}
    \theta(t) = -\frac{3\pi J_2 R^2_* \mathrm{cos}\psi}{Pa^2}t + \theta_0
\end{equation}

\noindent where $J_2$ is the stellar gravitational quadruple moment. The formulation assumes that the angular momentum due to stellar rotation is much larger than that due to the planetary orbit, causing the orbital axis to precess around the stellar rotational axis. Such assumption may not always hold for hot Jupiters with sufficient orbital angular momentum \citep{Barnes_2013}, although this does not significantly affect the precession rate.

We calculate $b(t)$ and $\lambda(t)$ for KELT-19Ab, assuming $J_2 = (1.36^{\mathrm{+0.15}}_\mathrm{-0.12})\times 10^{-4}$ taken from \cite{Watanabe_2022} in their analysis of WASP-33b, in which the host WASP-33 has a similar $T_\mathrm{eff} \approx 7430$ K and $v\mathrm{sin}i_* \approx 85.6$ km/s as KELT-19A. Under this assumption, impact parameter for the polar ($i_* \approx 34^\circ$) and flipped ($i_* \approx 97^\circ$) case will only deviate by around 0.04 and by about a degree for sky-projected obliquity in the next three decades, neither of which are deviations large enough to be detected.

\subsection{Possible scenarios}
\subsubsection{Kozai-Lidov mechanism}
Considering that a suitable companion exists in KELT-19B, which is a late-G9V/early-K1V star located at a projected separation of 160 au \citep{Siverd_2018}, Kozai-Lidov mechanism is a plausible hypothesis to explain the orbital evolution of KELT-19Ab. The timescale for Kozai-Lidov mechanism can be calculated using the following approximation. 

\begin{equation}
P_\mathrm{Kozai} \sim \frac{M_*}{M_c}\frac{P^2_c}{P}(1-e^2_c)^{3/2}    
\end{equation}

\noindent where $M_c$, $P_c$, and $e_c$ are the mass, the orbital period and the eccentricity of the companion. To estimate the this timescale for KELT-19Ab, we assume $0.9M_\odot$ for the late-G9V/early-K1V companion KELT-19B from \cite{mamajek}, and take the projected separation of the companion of 160 au from \cite{Siverd_2018} to calculate $P_c$. Considering a full range of orbital eccentricity of KELT-19B and accounting for the underestimation of the actual separation with respect to the projected separation, we obtain that the timescale does not exceed 0.2~Gyr, which is well below the estimated system age of 1.1 Gyr \citep{Siverd_2018}. Hence, the estimated timescale is consistent with the hypothesis that Kozai-Lidov mechanism played a role in KELT-19Ab's orbital evolution.

If KELT-19Ab's orbit is flipped and is beyond $140.8^\circ$ (known as the Kozai angle), however, traditional application of Kozai-Lidov mechanism \citep[e.g.]{Fabrycky_2007} fails to explain such large obliquity. This is because the libration of pericenter essential in inducing the oscillation of eccentricity and inclination only occurs for systems below this critical angle. This angle is therefore the upper limit of obliquity resulting from Kozai-Lidov mechanism (lower limit exists similarly for prograde planets at $39.2^\circ$), where obliquity in this context is defined to be the angle between the initial (pre-Kozai) and the evolved (post-Kozai) planetary orbital axis. Numerical simulations show that octupole level effects allow for some dispersion around these angles \citep{Petrovich_2015,Petrovich_2016}, but not to the extent that planets with $\psi \approx 180^\circ$ are expected. 

Meanwhile, obliquites beyond the Kozai angle can be achieved when the concurrent evolution of stellar spin is taken into account during Kozai-Lidov mechanism. \cite{Storch_Anderson_Lai_2014} found that chaotic evolution of stellar spin due to spin-orbit coupling between the precession frequencies of the oblate star and the Kozai frequencies result in extreme obliquities, some of them close to $180^\circ$. The result owes to the fact that obliquity in this context is no longer the angle between the pre and post-Kozai planetary orbital axis assuming primordial alignment, but rather the angle between the stellar spin axis and the planetary orbital axis. The likeliness of a star experiencing this chaotic spin evolution is quantified in \cite{Stroch_Lai_2015} by the adiabacity parameter $\epsilon$,

\begin{align}
\epsilon & =\left|\frac{\Omega_{\mathrm{pl}}}{\Omega_{\mathrm{ps}}}\right|_{e, \theta_{\mathrm{sl}}=0} \notag\\
& = \, 1.17\left(\frac{k_{\star}}{2 k_q}\right)\left(\frac{R_{\star}}{1 \mathrm{R}_{\odot}}\right)^{-3 / 2}\left(\frac{\hat{\Omega}_{\star}}{0.1}\right)^{-1}\left(\frac{M_{\mathrm{c}}}{10^3 M_{\mathrm{p}}}\right) \notag\\ & \times\left(\frac{a}{1 \mathrm{au}}\right)^{9 / 2}\left(\frac{a_c}{300 \mathrm{au}}\right)^{-3}\left|\cos \theta_{\mathrm{lc}}^0\right|,
\end{align}

\noindent where $\Omega_{\mathrm{pl}}$ and $\Omega_{\mathrm{ps}}$ are the precession rate of the planet and star around the companion respectively, and $\epsilon$ on the order of unity (i.e. $\Omega_{\mathrm{pl}} \approx \Omega_{\mathrm{ps}}$) is the required criterion for a chaotic evolution. Relevant parameters are the stellar moment of inertia constant $k_*$, 'rotational distortion constant' $k_q$ which is one third the Love number $k_2$, stellar raidus $R_*$, stellar rotation rate $\hat{\Omega}_{\star} = {\Omega}_{\star}/(GM_*/R^3_*)^{1/2}$, companion and planetary mass $M_c$ and $M_p$ as well as semi-major axis $a_c$ and $a$, and the initial mutual inclination between them $\theta^0_{\mathrm{lc}}$, which we use $140.8^\circ$ to calculate the upper bound of $\epsilon$. We also follow \cite{Stroch_Lai_2015} and adopt the canonical values $k_* \approx 0.1$ and $k_q \approx 0.05$ \citep{Claret_1992}. We obtain $\epsilon \approx 1.7$ for KELT-19Ab when semi-major axis of 1 au is assumed, but a gas giant like KELT-19Ab is unlikely to have formed at such a short distance considering 
the location of snowline \citep{Kennedy_2008}. The parameter is very sensitive to the assumed semi-major axis and quickly shoots up ($\epsilon \approx 5 \times 10^4$ when $a$ = 10 au), likely indicating that a chaotic evolution of stellar spin is not a suitable mechanism to explain the obliquity of KELT-19Ab.

Even if the direction of stellar spin remains unchanged for the duration of Kozai-Lidov mechanism, flipped orbits beyond the Kozai angle can also result from primordial misalignment between the stellar spin axis and the planetary orbital axis \citep{Anderson_2016, Vick_2022}. Misaligned inner disks \citep{Marino_2015,Ansdell_2020} provide strong observational evidence for primordial misalignments, and misaligned coplanar multi-planet systems  \citep{Huber_2013,Hjorth_2021} are potentially explained by such misalignmet as well. 

Theoretically, gravitational interactions between host stars, protoplanetary disks, and inclined binary companions are a way to form primordial misalignments \citep{Batygin_2013,Lai_2014,Spalding_2014,Zanazzi_2018}. \cite{Zanazzi_2018} showed, however, that when taken into account spin–orbit coupling between the host star and the planet as well as planet–disc interactions, excitation of primordial misalignment can either be completely suppressed or at best significantly reduced for hot Jupiters whether formed in situ or migrated inwards via type-II migration. There are nonetheless several other ways to form primordially misaligned planets including the nonaxisymmetric collapse of the molecular cloud core \citep{Bate_2010, Takaishi_2020}, magnetic star–disc interactions \citep{Lai_Dong_Foucart_2011}, or dissipation of internal-gravity waves in the host star \citep{Rogers_2013}, which when combined with Kozai-Lidov mechanism may explain the obliquity of KELT-19Ab.

\subsubsection{Other mechanisms}
Coplanar high-eccentricity migration \citep{Li_2014, Petrovich_2015} is another theory predicting the formation of planets with $\psi \approx 180^\circ$, where secular gravitational interactions between two coplanar and eccentric planets cause their orbits to potentially flip. In such a case, even when the orbit of a planet initially begins prograde with respect to the orbit of the distant companion, the secular interaction causes the orbit to completely flip. For such an orbital flip to occur, the semi-major axis of the inner planet must be large enough to avoid being tidally disrupted from the extreme eccentricity growth preceding the flip, which is hardly achieved \citep{Petrovich_2015,Xue_2016}.\footnote{\label{foot:flip}Orbital flip (an initially prograde planet becomes retrograde or vice versa) itself is possible with Kozai-Lidov mechanism between two planets, or even with a stellar companion when taking into account ocutuple order effects, but again not to the extent that planets with $\psi \approx 180^\circ$ are expected \citep{Naoz_2011}. }

\cite{Xue_2014} have also suggested that tidal dissipation of inertial waves in the convective envelope of the star can temporarily realign the planet's orbit to $\psi \approx 180^\circ$. However, KELT-19A with the $T_\mathrm{eff} = 7500\pm110$ K is expected to have a radiative envelope rather than a convective one, and tidal realignment is improbable. While limited to systems in a very dense stellar cluster, another proposed mechanism fly-bys of a prograde and equal-mass perturber \citep{Breslau_2019}.

\subsection{Implications on the inferred obliquity distribution}
The large obliquity of KELT-19Ab is consistent with the recent finding due to Hierarchical Bayesian Analysis that there is no statistically significant preference towards polar planets, and that obliquity can be distributed more uniformly \citep{Siegel_2023,Dong_2023}. These studies refuted the frequentist test that showed that the observed distribution of misaligned planets (defined as $\mathrm{cos}\psi < 0.75 \,\, \mathrm{or}\,\, \psi > 41^\circ$) is significantly different from a uniform distribution between $41^\circ < \psi < 180^\circ$ \citep{Albrecht_2021}.

For the purpose of Hierarchical Bayesian Analysis, both \cite{Siegel_2023} and \cite{Dong_2023} concluded that measurements of $\lambda$ was already sufficient to establish the uniformity of the inferred distribution. $\lambda$ measured in \cite{Siverd_2018} was already incorporated in both studies, and we therefore decide not to replicate the the analysis with our constraints on $\psi$. The ability of $\lambda$ in the context of Hierarchical Bayesian Analysis, nevertheless, does not undermine the importance of measuring $\psi$ for individual systems. Especially when flipped orbits are in interest, large $\lambda$ does not guarantee a flipped orbit \citep{Fabrycky_Winn_2009,Xue_2016}. Then, measuring $\lambda$ is insufficient, unlike when we have $\lambda \approx \psi$ regardless of $i_*$, where $\lambda \approx 90^\circ$. Therefore, measurements of $\psi$ is indispensable when searching for flipped orbits.

\section{Conclusion} \label{sec:cite}
In this work, KELT-19Ab's symmetric light curve from TESS was analyzed with gravity darkening to estimate its obliquity. We find that with constraints on stellar rotational velocity, sky-projected obliquity as well as limb-darkening coefficients, KELT-19Ab's obliquity is $\psi = 155^{\circ+17^\circ}_{\,\,-21^\circ}$ ($119^\circ<\psi<177^\circ$ at $2\sigma$), in favor of a flipped orbit. The finding is consistent with the uniformity of the inferred obliquity distribution presented in both \cite{Siegel_2023} and \cite{Dong_2023} using Hierarchical Bayesian Analysis.

From theoretical standpoints, formations of flipped orbits are rare but possible with Kozai-Lidov mechanism even for the most extreme cases, when accounting for the concurrent evolution of stellar spin \citep[e.g.]{Storch_Anderson_Lai_2014,Stroch_Lai_2015} or primordial misalignments between the planetary orbital axis and stellar spin axis \citep{Anderson_2016,Vick_2022}. For KELT-19Ab, the latter case of Kozai-Lidov mechanism coupled with primordial misalignments is a plausible hypothesis.

We also investigate the previously discussed but unresolved discrepancy in transit depth, impact parameter and semi-major axis which might be explained by the gravity darkened transit of KELT-19Ab. Although the trend of larger impact parameters coupled with larger transit depth precisely matches the characteristic of degeneracy caused by gravity darkened transits of planets with $\lambda$ and $\psi$ both nearing $180^\circ$ \citep{Barnes_2009}, the results were inconclusive as to whether gravity darkening was the lone culprit of the discrepancies in KELT-19Ab. Similar discrepancies, however, if more pronounced might be another way to tell a flipped orbit from a polar one in other systems.

The above results highlight the applicability of the gravity darkening method to symmetric light curves, which was previously set aside the visibly asymmetric ones. As discussed by \cite{Barnes_2009}, it is nonetheless crucial that a constraint on stellar rotational velocity is given apriori to gauge the degree of gravity darkening. Constraints on sky-projected obliquity is also vital to limit the possible shapes of gravity darkened light curves, and on limb-darkening coefficients to reduce the degeneracy in gravity darkening.

\section*{Data Availability}
The data used in this paper is publicly available from the Mikulski Archive for Space Telescopes (MAST) and produced by the Science Processing Operations Center (SPOC) at NASA Ames Research Center \citep{Jenkins_2016,Jenkins_2017}. Systematics reduced light curves are obtained via the subtraction of cotrending basis vectors (CBVs) using the Python package $\mathrm{\begin{tt}eleanor\end{tt}}$ \citep{eleanor}.

\section*{Acknowledgement}
We would like to thank the referee, Dr. Jason W. Barnes for his insightful suggestions. YK would like to thank Omar Attia, Jan-Vincent Harre, Kento Masuda and J.J. Zanazzi for their insights and comments on this work at PPVII. Work by YK is funded by the University of Tokyo, World-leading Innovative Graduate Study Program of Advanced Basic Science Course. This work is partly supported by JSPS KAKENHI Grant Numbers JP18H05439, JP21K20376, and JST CREST Grant Number JPMJCR1761. This paper includes data collected by the TESS mission. Funding for the TESS mission is provided by the NASA's Science Mission Directorate.



\bibliographystyle{mnras}
\bibliography{KELT_19Ab_paper} 





\appendix
\section{Corner plots}
\label{corner_plots}
We display the corner plots obtained with the two models below.
\begin{figure*}
    \centering
    \includegraphics[width=2\columnwidth]{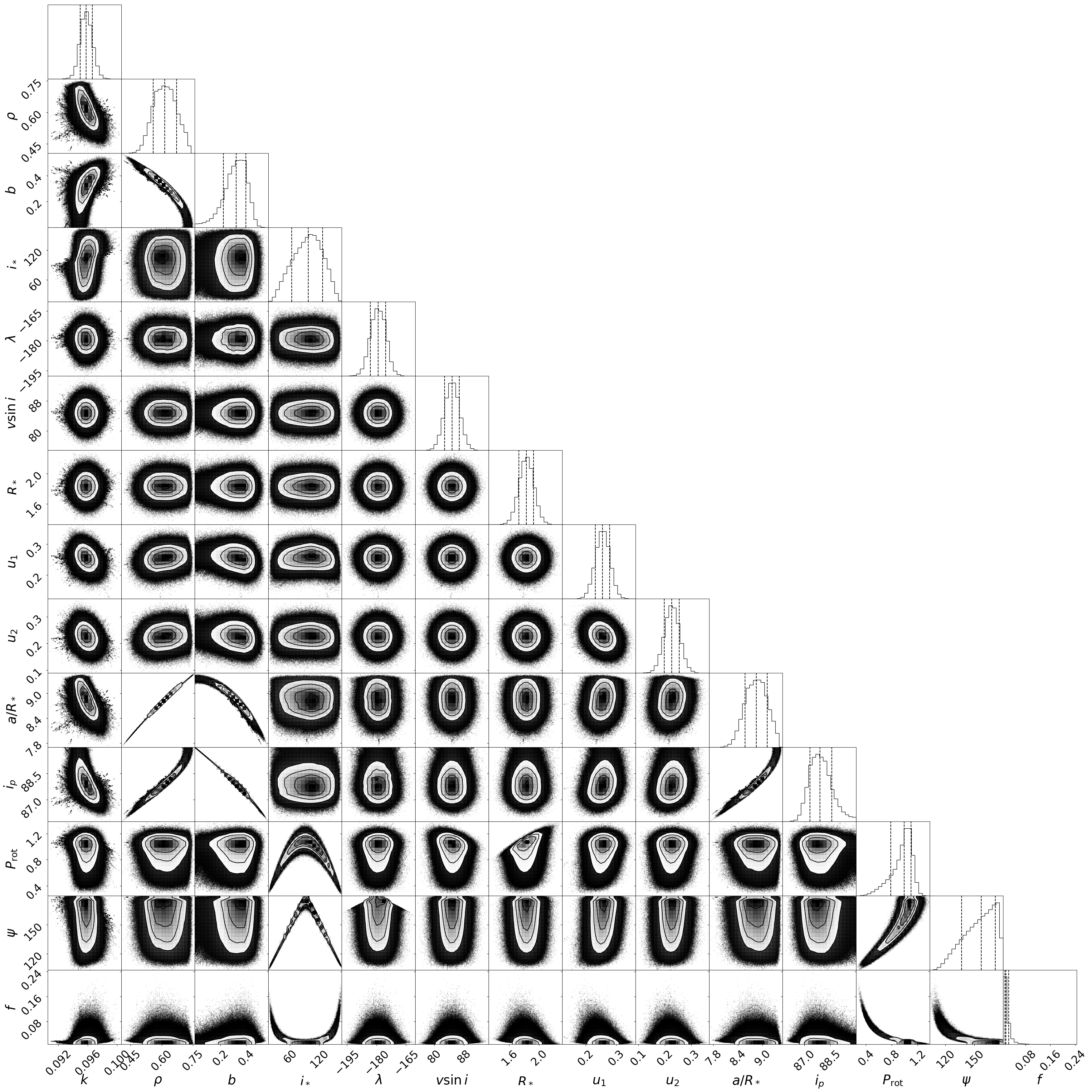}
    \caption{Corner plot for GD7400.}
    \label{fig:corner_plot_7400}
\end{figure*}
\begin{figure*}
    \centering
    \includegraphics[width=2\columnwidth]{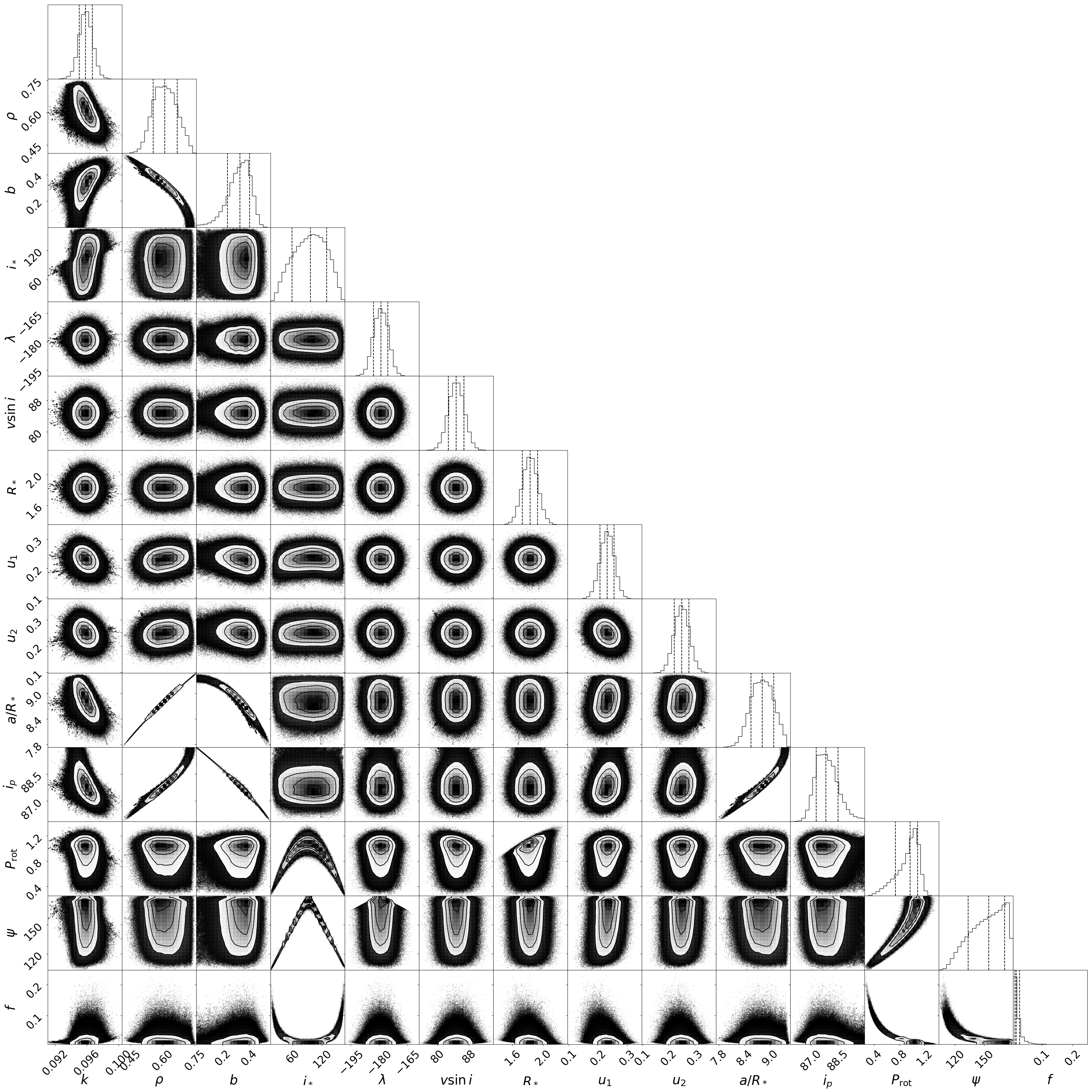}
    \caption{Corner plot for GD7600.}
    \label{fig:corner_plot_7600}
\end{figure*}
\bsp	
\label{lastpage}
\end{document}